\documentclass[12pt,a4paper]{article}
\usepackage{graphicx}
\newcommand{\vs}{\vspace{-0.25cm}}

\begin{document}

\begin{center}

{\Large
\textbf{Nuclear energy density functional from chiral\\ pion-nucleon 
dynamics: Isovector terms}}

\bigskip
N. Kaiser\\

\bigskip

{\small Physik Department T39, Technische Universit\"{a}t M\"{u}nchen, 
D-85747 Garching, Germany\\

\smallskip

{\it email: nkaiser@ph.tum.de}}

\end{center}

\bigskip

\begin{abstract}
We extend a recent calculation of the nuclear energy density functional in the 
framework of chiral perturbation theory by computing the isovector
surface and spin-orbit terms: $(\vec \nabla  \rho_p- \vec \nabla \rho_n)^2 \, 
G_d(\rho)+(\vec \nabla \rho_p- \vec \nabla \rho_n)\cdot(\vec J_p-\vec J_n)\,
G_{so}(\rho)+(\vec J_p-\vec J_n)^2\,G_J(\rho)$ pertaining to different proton and
neutron densities. Our calculation treats systematically the effects from 
$1\pi$-exchange, iterated $1\pi$-exchange, and irreducible $2\pi$-exchange with
intermediate $\Delta$-isobar excitations, including Pauli-blocking corrections 
up to three-loop order. Using an improved density-matrix expansion, we obtain 
results for the strength functions $G_d(\rho)$, $G_{so}(\rho)$ and $G_J(\rho)$ 
which are considerably larger than those of phenomenological Skyrme forces. 
These (parameter-free) predictions for the strength of the isovector surface 
and spin-orbit terms as provided by the long-range pion-exchange dynamics 
in the nuclear medium should be examined in nuclear structure calculations at 
large neutron excess.  
\end{abstract}

\bigskip


\bigskip 

\section{Introduction}
The nuclear energy density functional approach is the many-body method of 
choice in order to calculate the properties of medium-mass and heavy nuclei in 
a systematic manner \cite{reinhard}. Parameterized (non-relativistic) Skyrme 
functionals \cite{sk3,skmstar,skp,sly,pearson} as well as relativistic
mean-field models \cite{walecka,ringreview} have been used widely and
successfully for such nuclear structure calculations. Likewise, some 
constraints from chiral (pion-nucleon) dynamics and the symmetry breaking 
pattern of QCD at low energies have been implemented into a pertinent 
relativistic point-coupling Lagrangian in ref.\cite{finelli}. 

A complementary approach in the quest for a predictive nuclear energy density
functional \cite{lesinski,drut,platter} focusses less on the fitting of 
experimental data, but attempts to constrain the analytical form of the 
functional and the values of its couplings from many-body perturbation theory
and the underlying two- and three-nucleon interactions. Switching from the
conventional hard-core NN-potentials to low-momentum interactions is essential
in this respect, because the nuclear many-body problem formulated in terms of 
the latter becomes significantly more perturbative. Indeed, second-order 
perturbative calculations provide already a good account of the bulk 
correlations in infinite nuclear matter \cite{achim} and in doubly-magic 
nuclei \cite{roth}. 

In many-body perturbation theory the contributions to the energy are written 
in terms of density-matrices and propagators convoluted with the finite-range 
interaction vertices, and are therefore highly non-local in both space and 
time. In order to make such functionals numerically tractable in heavy 
open-shell nuclei it is desirable to develop simplified approximations to these 
functionals expressed in terms of local densities and currents only. In this 
construction the density-matrix expansion comes prominently into play as it 
removes the non-local character of the exchange (Fock) contribution to the
energy by mapping it onto a generalized Skyrme functional with 
density-dependent couplings. For some time the prototype for that has been 
the density-matrix expansion of Negele and Vautherin \cite{negele}, but recently
Gebremariam, Duguet and Bogner \cite{dmeimprov} have developed an improved 
version for spin-unsaturated nuclei. They have demonstrated that 
phase-space averaging techniques allow for a consistent expansion of both the 
spin-independent (scalar) part as well as the spin-dependent (vector) part of 
the density-matrix. The accuracy of the new phase-space averaged density-matrix
 expansion and the original one of Negele and Vautherin \cite{negele} has been 
gauged via the Fock energy (densities) arising from (schematic finite-range) 
central, tensor and spin-orbit interactions for a large set of semi-magic 
nuclei. For a central force the Fock energy depends primarily on the 
spin-independent (scalar) part of the  density-matrix and a few percent 
accuracy is reached for both variants of the density-matrix expansion. On the 
other hand the Fock energy due to a tensor force is determined by the 
spin-dependent (vector) part of the density-matrix. In that case the original 
density-matrix expansion of Negele and Vautherin \cite{negele} leads to an 
error of about $50\%$, whereas the new one based on phase-space averaging
techniques reduces the error drastically to only a few percent. For further
details on these extensive and instructive test studies we refer to
ref.\cite{dmeimprov}. 
          
In order to match with these new developments the nuclear energy density 
functional as it emerges from chiral pion-nucleon dynamics has been recalculated
recently in ref.\cite{energfun}. That calculation has treated for
isospin-symmetric  (i.e. $N=Z$) nuclear systems  the effects from 
$1\pi$-exchange, iterated $1\pi$-exchange, and irreducible $2\pi$-exchange 
with intermediate $\Delta$-isobar excitations, including Pauli-blocking
corrections up to three-loop order. It has been found that the effective
nucleon mass $M^*(\rho)$ entering the energy density functional becomes
identical to the one of Fermi-liquid theory when employing the improved
density-matrix expansion. The strength $F_\nabla(\rho)$ of the $(\vec\nabla 
\rho)^2$  surface term as provided by the pion-exchange dynamics comes out in 
good agreement with empirical values in the density region $\rho_0/2 <\rho <
\rho_0=0.16\,{\rm fm}^{-3}$. The spin-orbit coupling strength $F_{so}(\rho)$ 
receives contributions from iterated $1\pi$-exchange (of the ''wrong sign'')
and from three-nucleon interactions mediated by $2\pi$-exchange with virtual 
$\Delta$-excitation (of the ''correct sign''). In the region around $\rho_0/2 
\simeq 0.08\,$fm$^{-3}$ where the spin-orbit interaction in nuclei gains most
of its weight these two components tend to cancel, thus leaving all room for 
the short-range spin-orbit interaction. As a matter of fact the empirical 
value $F_{so}^{(\rm  emp)}\simeq 90\,$MeVfm$^5$ of the spin-orbit coupling strength 
in nuclei agrees well with the one extracted from the short-range spin-orbit 
component of any realistic NN-potential \cite{short}. This part of the 
NN-interaction drives at the same time the strong Lorentz scalar and vector 
mean-fields \cite{tueb} on which the whole success of the relativistic Dirac 
phenomenology rests.

The purpose of the present paper is the extend the calculation of the energy 
density functional in ref.\cite{energfun} to isospin-asymmetric many-nucleon 
systems with different proton and neutron densities. The supplementary 
isovector surface and spin-orbit terms play an important role in the
description of long chains of stable isotopes and for nuclei far from 
stability. Our paper is organized as follows. In section 2, we recall the 
improved density matrix expansion of Gebremariam, Duguet and Bogner 
\cite{dmeimprov} whose Fourier transform to momentum space  provides the 
adequate technical tool to calculate the nuclear energy density functional in 
a diagrammatic framework. In section 3, we present the analytical results for 
the density-dependent strength functions $G_d(\rho)$, $G_{so}(\rho)$ and 
$G_J(\rho)$ belonging to the isovector surface and spin-orbit terms. These 
analytical expressions give individually the effects due to  $1\pi$-exchange, 
iterated $1\pi$-exchange, and irreducible $2\pi$-exchange with intermediate 
$\Delta$-isobar excitations, including Pauli-blocking corrections up to  
three-loop order. Section 4 is devoted to a discussion of our numerical 
results and it ends with some concluding remarks and an outlook.

\section{Energy density functional and improved density-matrix expansion}
Let us begin with writing down the explicit form of the isovector surface and
spin-orbit terms in the nuclear energy density functional: 
\begin{equation} {\cal E}_{\rm iv}[\rho_p,\rho_n,\vec J_p,\vec J_n] =(\vec  \nabla  
\rho_p- \vec \nabla \rho_n)^2 \, G_d(\rho) + (\vec \nabla \rho_p- \vec \nabla 
\rho_n) \cdot(\vec J_p-\vec J_n)\,G_{so}(\rho)+(\vec J_p-\vec J_n)^2\,G_J(\rho)\,. 
\end{equation} 
Here, 
\begin{equation} \rho_{p,n}(\vec r\,) = {k_{p,n}^3(\vec r\,) \over 3\pi^2} = 
\sum_{\alpha } \Psi^{(\alpha)
\dagger}_{p,n}( \vec r\,)\Psi^{(\alpha)}_{p,n}( \vec r\,)\,,\end{equation} 
are the local proton and neutron densities written in terms of the 
corresponding (local) proton and neutron Fermi momenta $k_{p,n}(\vec r\,)$, and
expressed as sums over the occupied single-particle orbitals $\Psi^{(\alpha)}_{p,n}
(\vec r\,)$. The spin-orbit densities of protons and neutrons are 
defined similarly: 
\begin{equation} \vec J_{p,n}(\vec r\,) = \sum_{\alpha} \Psi^{(\alpha)
\dagger}_{p,n}(\vec r\,)i\, \vec \sigma \times \vec \nabla\Psi^{(\alpha)}_{p,n
}( \vec r\,) \,. \end{equation} 
Furthermore, $G_d(\rho)$, $G_{so}(\rho)$ and $G_J(\rho)$ in eq.(1) denote the 
associated strength functions depending on the total nucleon density 
$\rho=\rho_p+\rho_n$. In Skyrme parameterizations 
\cite{sk3,skmstar,skp,sly,pearson} these are just constants, $G_d^{(\rm Sk)} = 
-[3t_1(2x_1+1)+t_2(2x_2+1)]/64$, $G_{so}^{(\rm  Sk)}=W_0/4$, $G_J^{(\rm Sk)}=
(t_1-t_2)/32$, whereas in our calculation their explicit density-dependence 
originates from the finite-range character of the $1\pi$- and $2\pi$-exchange 
interaction.

The starting point for the construction of an explicit nuclear energy density 
functional ${\cal E}_{\rm iv}[\rho_p,\rho_n,\vec J_p,\vec J_n]$ is the bilocal 
density-matrix as given by a sum over the occupied energy eigenfunctions: 
$\sum_{\alpha}\Psi^{(\alpha)}_{p,n}( \vec r -\vec a/2)\Psi^{(\alpha)\dagger}_{p,n}(\vec r +
\vec a/2)$. According to  Gebremariam, Duguet and Bogner \cite{dmeimprov} it 
can be expanded in relative and center-of-mass coordinates, $\vec a$ and 
$\vec r$, with expansion coefficients determined by local quantities (nucleon 
density, kinetic energy density and spin-orbit density). As outlined in 
section 2 of ref.\cite{energfun} the Fourier transform of the expanded 
density-matrix with respect to both coordinates $\vec a$ and $\vec r$ defines 
in momentum space a ''medium-insertion'' $\Gamma(\vec p,\vec q\,)$ for the 
inhomogeneous many-nucleon system. It is straightforward to generalize this 
construction \cite{energfun} to the isospin-asymmetric situation with different 
proton and  neutron local densities $\rho_{p,n}(\vec r\,)$ and $\vec J_{p,n}(\vec 
r\,)$. We display here only that part of the medium-insertion $\Gamma(\vec p,
\vec q\,)$ which is actually relevant for the diagrammatic calculation of the 
isovector  surface and spin-orbit terms introduced in eq.(1): 
\begin{eqnarray} \Gamma(\vec p,\vec q\,)& =& \int d^3 r \, e^{-i \vec q \cdot
\vec r}\,\bigg\{ {1+\tau_3 \over 2}\,\theta(k_p-|\vec p\,|) +{1-\tau_3 \over 2}
\,\theta(k_n-|\vec p\,|) \nonumber \\ && -{3\pi^2 \over 4k_f^4}
\delta(k_f-|\vec p\,|)\, \tau_3 \, \vec \sigma\cdot [\vec p \times 
(\vec J_p-\vec J_n)] +\dots  \bigg\}\,.  \end{eqnarray}
When working to quadratic order in deviations from isospin symmetry
(proton-neutron differences) it is sufficient to use an average Fermi momentum 
$k_f$ in the prefactor of the isovector spin-orbit density $\vec J_p-\vec J_n$.
The double line in the left picture of Fig.\,1 symbolizes this medium insertion
together with the assignment of the out- and in-going nucleon momenta $\vec p 
\pm \vec q/2$. The momentum transfer $\vec q$ is provided by the Fourier 
components of the inhomogeneous (matter) distributions $\rho_{p,n}(\vec r\,)$ 
and $\vec J_{p,n}(\vec r\,)$. Note that in comparison to the version of 
$\Gamma(\vec p,\vec q\,)$ which followed from Negele and Vautherin's 
density-matrix expansion \cite{negele} the weight function of the spin-orbit 
densities $\vec J_{p,n}(\vec r\,)$ has changed from  $\delta(k_f-|\vec p\,|) 
-k_f\,\delta'(k_f-|\vec p\,|)$ to $-3 \delta(k_f-|\vec p\,|)$. Furthermore, a
pairwise filling of time-reversed orbitals $\alpha$ for both protons and
neutrons has been assumed, so that (various possible) time-reversal-odd fields
do not come into play  \cite{reinhard}.

\begin{figure}
\begin{center}
\includegraphics[scale=1.0,clip]{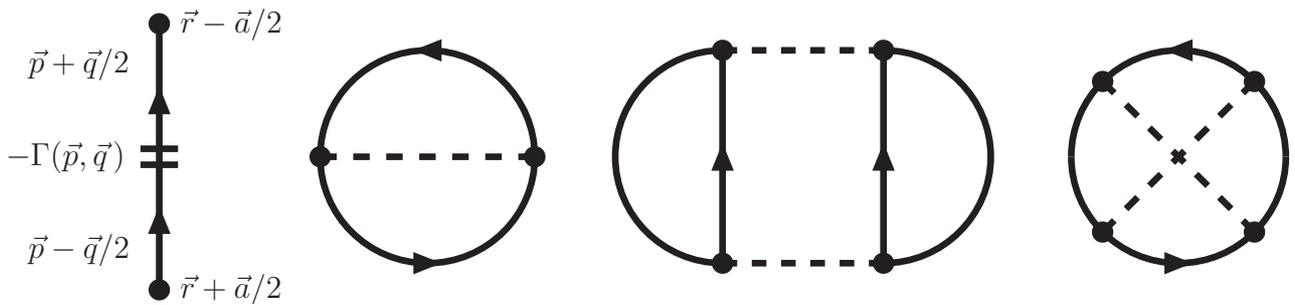}
\end{center}
\vspace{-0.2cm}
\caption{Left: The double-line symbolizes the medium insertion defined by
eq.(4). Next are shown: The one-pion exchange Fock diagram and the iterated 
one-pion exchange Hartree and Fock diagrams.}
\end{figure}

\section{Diagrammatic calculation}
In this section we present analytical formulas for the three density-dependent
strength functions $G_d(\rho)$, $G_{so}(\rho)$ and $G_J(\rho)$ as derived (via 
the improved density-matrix expansion \cite{dmeimprov}) from $1\pi$-exchange, 
iterated $1\pi$-exchange, and irreducible $2\pi$-exchange diagrams with 
intermediate $\Delta$-isobar excitations, including Pauli-blocking corrections 
up to three-loop order. We give for each diagram only the final result 
omitting all technical details related to extensive algebraic manipulations 
and solving elementary integrals. In essence the calculation of 
ref.\cite{energfun} gets just modified by relative isospin factors $-1/3,\, 
\pm 2/3,\, -5/3$ occurring at various places.   

\subsection{One-pion exchange Fock diagram}
The non-vanishing contribution from the $1\pi$-exchange Fock diagram shown in 
Fig.\,1 reads:
\begin{equation} G_J(\rho) = {3g_A^2 \over(32m_\pi f_\pi)^2 u^6}\Big[4u^2-8u^4 
-\ln(1+4u^2) \Big] \,,\end{equation}
where we have introduced the convenient dimensionless variable $u =k_f/m_\pi$. 
This contribution to $G_J(\rho)$ is just $-1/3$ of the contribution to the
isoscalar strength function $F_J(\rho)$ (see eq.(11) in ref.\cite{energfun}) as
a  consequence of the isospin trace: tr$[\tau_a (J_s+\tau_3 J_v)\tau_a(J_s+\tau_3
 J_v)]= 6 J_s^2 - 2J_v^2$.
\subsection{Iterated $1\pi$-exchange Hartree diagram with two medium
  insertions}
The two-body contributions from the iterated $1\pi$-exchange Hartree diagram
in Fig.\,1 read: 
\begin{equation} G_d(\rho) = {g_A^4M \over 3\pi m_\pi(4f_\pi)^4} \bigg\{
{23 \over 8u^2} \ln(1+4u^2)-{8\over u} \arctan 2u  +{3+20u^2\over 6(1+4u^2)^2}  
\bigg\} \,,  \end{equation}
\begin{equation} G_{so}(\rho) =  {g_A^4 M \over \pi m_\pi (4f_\pi u)^4}\bigg\{ 
6u^2+{5\over 2}\ln(1+4u^2)-8u \arctan 2u \bigg\} \,, \end{equation}
which are $-2/3$ of the respective isoscalar contributions 
\cite{energfun,efun}. Let us briefly explain the mechanism which generates the 
strength function $G_d(\rho)$. The exchanged pion-pair transfers the momentum 
$\vec q$ between the left and the right nucleon ring. This momentum $\vec q$ 
enters both the pseudovector $\pi N$-interaction vertices and the pion 
propagators. After expanding the inner loop integral to order $\vec q\,^2$ the 
Fourier transformation in eq.(4) converts this factor $\vec q\,^2$ into a 
factor $(\vec \nabla k_p-\vec \nabla k_n)^2 \simeq \pi^4(\vec \nabla \rho_p- 
\vec \nabla \rho_n)^2/k_f^4$. The rest is a solvable integral over the product
of two Fermi spheres of radius $k_f$. The isovector spin-orbit strength
function $G_{so}(\rho)$ arises from the spin-trace: tr$[\vec \sigma\cdot(\vec 
l+\vec q/2)\,\vec \sigma\cdot (\vec l -\vec q/2)\,\vec \sigma \cdot (\vec 
p_{1,2} \times \vec J_v)] = 2i\,(\vec q \times \vec l\,)\cdot(\vec p_{1,2} \times 
\vec J_v)$, where $\vec q$ gets again converted to $\vec \nabla k_p-\vec \nabla
k_n \simeq \pi^2(\vec \nabla \rho_p- \vec \nabla \rho_n)/k_f^2$ by Fourier 
transformation. 
\subsection{Iterated $1\pi$-exchange Fock diagram with two medium insertions}
We find the following contributions from the right Fock diagram in Fig.\,1 
with two medium insertions on non-neighboring nucleon propagators:
\begin{eqnarray} G_d(\rho) &=& {5g_A^4M \over 3\pi m_\pi(8f_\pi)^4} \bigg\{
{4\over u}(2 \arctan 2u- \arctan u)+{1\over u^2}\ln{(1+u^2)^2 \over (1+2u^2)
(1+4u^2) } \nonumber \\ &&-{4\over 1+2u^2} +{2\over u^2}\int_0^u \!dx\,
{3+18x^2+16x^4\over (1+2x^2)^3}\Big[\arctan x-\arctan 2x\Big] \bigg\} \,,  
\end{eqnarray}
\begin{eqnarray}  G_{so}(\rho) &=& {5g_A^4 M \over 2\pi m_\pi (4f_\pi u)^4}\bigg\{ 
-u^2+ \int_0^u \!\!dx \, {1\over 1+2x^2}  \nonumber \\ && \times
\Big[(1+4x^2) \arctan 2x -4x^2(1+x^2) \arctan x \Big] \bigg\}\,, \end{eqnarray}
\begin{eqnarray}  G_J(\rho) &=& {15g_A^4 M \over \pi m_\pi (8f_\pi u)^4}\bigg\{ 
u^2+{2\over u^2} \int_0^u \!\!dx \, {1\over 1+2x^2} \Big[(x^2-u^2)(1+4x^2) 
\arctan 2x \nonumber \\ &&+2(6x^6-2u^2x^4+6x^4-2u^2x^2+2x^2-u^2)\arctan x \Big]
\bigg\}\,. \end{eqnarray}
One notices a relative isospin factor of $-5/3$ in comparison to the respective
 isoscalar contributions \cite{energfun,efun} which comes from the isospin 
trace tr$[\tau_a (J_s+\tau_3 J_v) \tau_b\tau_a (J_s+\tau_3J_v)\tau_b]= 10 J_v^2 - 
6J_s^2$ of this diagram.  
\subsection{Iterated $1\pi$-exchange Hartree diagram with three medium insertions}
In our way of organizing the many-body calculation, the Pauli-blocking 
corrections are represented by diagrams with three medium insertions. The
corresponding contributions from the iterated $1\pi$-exchange Hartree diagram 
shown in Fig.\,1 read: 
\begin{eqnarray}  G_d(\rho) &=& {2g_A^4 M \over 3\pi^2 m_\pi (4f_\pi)^4} \Bigg\{
{1\over u}\int_0^1 \!\! dy\,\ln{1+y \over 1-y} \bigg[{4u^2y^2(6+51u^2y^2+92 
u^4y^4) \over 3(1+4u^2y^2)^3}  \nonumber \\ && -2 \ln(1+4u^2y^2)\bigg]
+ \int_0^u \!\! dx\,{x^2\over u^4} \int_{-1}^1 \!\! dy\,{s^3  s'(3+9s^2+4s^4) 
\over (1+s^2)^4} \ln{u+x y \over u-x y} \Bigg\}\,, \end{eqnarray}  
\begin{eqnarray}  G_{so}(\rho) &=& {2g_A^4 M \over \pi^2 m_\pi (4f_\pi)^4}
\int_0^u \!\! dx\,{x^2\over u^6} \int_{-1}^1 \!\! dy\,\Bigg\{\bigg[ 4x y 
\ln{u+x y \over u-x y} +{u(5x^2y^2-3u^2) \over  u^2-x^2y^2}\bigg]\nonumber \\ &&
\times \bigg[5s+{s\over (1+s^2)^2} -6 \arctan s\bigg] -{u s^5 (u^2+x^2y^2) 
\over (1+s^2)^2 (u^2-x^2y^2)}\nonumber \\ && +{2s^4 s'\over (1+s^2)^2} 
\bigg[ 4u-(s+2xy) \ln{u+x y \over u-x  y}\bigg] \Bigg\}\,, \end{eqnarray}  
\begin{equation} G_J(\rho)={3g_A^4 M u^3\over 16 \pi^2 m_\pi f_\pi^4}\int_0^1\!\!dy 
\,{y^6 \over(1+4u^2y^2)^2} \bigg[- 2y+(y^2-1) \ln{1+y\over 1-y}\bigg] \,.
\end{equation}
with the auxiliary function $s= x y +\sqrt{u^2-x^2+x^2y^2}$ and its partial
derivative $s' = u \,\partial s/\partial u$. 
\subsection{Iterated $1\pi$-exchange Fock diagram with three medium insertions}
The evaluation of this diagram is most tedious. It is advisable to split the
contributions to the strength functions $G_d(\rho)$, $G_{so}(\rho)$ and 
$G_J(\rho)$ into ''factorizable'' and ''non-factorizable'' parts. These two 
pieces are distinguished by the feature of whether the nucleon propagator in
the denominator can be canceled or not by terms from the product of 
$\pi N$-interaction vertices in the numerator. We find the following 
''factorizable'' contributions:
\begin{eqnarray} G_d(\rho) &=& {g_A^4 M\over 6\pi^2 m_\pi (4f_\pi)^4} \Bigg\{ 
{95\arctan 2u \over 8(1+u^2)}-{1+6u^2 \over 32u^5} \ln^2(1+4u^2)\nonumber \\ && 
+{9+100u^2+184u^4-192u^6\over 16u^3(1+u^2)(1+4u^2)} \ln(1+4u^2) -{7u \over
1+4u^2}-{7 \over 4u} \nonumber \\ &&+{5\over u^2} \int_0^u \!dx\,\bigg\{-{L^2 
\over x^2}(1+u^2)(3+3u^2+x^2)-{3u^2\over x^2}+L\bigg[4u+ {6 u \over x^2}(1+u^2)
\nonumber \\ && +{x-4u \over 1+(u+x)^2}-{x+4u \over 1+(u-x)^2}+{2x\over 
[1+(u+x)^2]^2}-{2x\over [1+(u-x)^2]^2}\bigg] \bigg\} \Bigg\} \,,\end{eqnarray}
\begin{eqnarray} G_{so}(\rho) &=&  {g_A^4 M \over \pi^2 m_\pi (8f_\pi u)^4}\Bigg\{ 
4 \Big[19u^2-\ln(1+4u^2)\Big]\arctan 2u -60u^3-8u -{3\over u}\nonumber \\ &&+  
{3+14u^2+2u^4 \over 2u^3}\ln(1+4u^2)-{3+20u^2+16u^4 \over 16u^5}\ln^2(1+4u^2)
\nonumber \\ && +20\int_0^u \!\!dx\, \bigg\{L^2\Big[u^4-2u^2-3-3x^{-2}(1+u^2)^3
+(3+7u^2)x^2-5x^4\Big] \nonumber \\ && +6 u x^{-2}(1+u^2)^2 L -3u^2x^{-2}(1+u^2)
\bigg\} \Bigg\} \,,\end{eqnarray} 
\begin{eqnarray}  G_J(\rho) &=&  {3g_A^4 M \over \pi^2 m_\pi (8f_\pi u)^4}\Bigg\{ 
{185 u \over 4}-4u^3 -35\arctan 2u +{1+4u^2\over 16u^5}\ln^2(1+4u^2) \nonumber 
\\ && +{1\over u} +{99u^2-8-68u^4 \over 16u^3}\ln(1+4u^2) +5\int_0^u \!\!dx\,  
\bigg\{{L^2 \over u^2}\bigg[ -{3\over 2x^2}(1+u^2)^4 \nonumber \\ && +2(u^4-1)
(1+u^2) -(5+2u^2+5u^4)x^2 +(6+10u^2)x^4 -{11 x^6 \over 2}\bigg] \nonumber \\ &&
+{L\over u}\bigg[{3\over x^2}(1+u^2)^3-3u^4-2u^2+1\bigg] -{3\over 2x^2}(1+u^2)^2 
\bigg\} \Bigg\} \,.\end{eqnarray} 
with the logarithmic function:
\begin{equation} L(x,u)= {1\over 4x} \ln{1+(u+x)^2\over 1+(u-x)^2} \,.
\end{equation}
The ''non-factorizable'' contributions (stemming from nine-dimensional principal
value integrals over the product of three Fermi-spheres of radius $k_f$) read
on the other hand: 
\begin{eqnarray} G_d(\rho) &=& {g_A^4 M \over 3\pi^2 m_\pi(4f_\pi)^4} \int_{-1}^1
\!\!dy \int_{-1}^1 \!\!dz\, {yz \,\theta(y^2+z^2-1) \over  |yz|\sqrt{y^2+z^2-1}}
\bigg\{{2u y^2\,\theta(y)\theta(z)\over 1+4u^2y^2}\nonumber \\ && \times \bigg[
{2u^2z^2(3+4u^2z^2) \over (1+4u^2z^2)^2}-\ln(1+4u^2z^2)\bigg]+ \int_0^u\!\! dx\,
{5x^2ss't^3t'(1-5s^2-2s^4)\over 4u^6(1+s^2)^3(1+t^2)}\bigg\}\,,\end{eqnarray}
\begin{eqnarray} G_{so}(\rho) &=& {g_A^4 M \over \pi^2 m_\pi(4f_\pi)^4} \int_{-1}^1
\!\!dy \int_{-1}^1 \!\!dz\, {yz \,\theta(y^2+z^2-1) \over  |yz|\sqrt{y^2+z^2-1}}
\bigg\{{8 y^2 z\, \theta(y)\theta(z)\over 1+4u^2y^2}\nonumber \\ && \times  
\Big[2uz-\arctan(2uz )\Big]+ \int_0^u \!\! dx\,{5x^2 s^2s't^2t'\over 2u^8(1+s^2)
(1+t^2)}(s t+s x z-t x y) \bigg\}\,, \end{eqnarray}
\begin{eqnarray} G_J(\rho) &=& {3g_A^4 M \over \pi^2 m_\pi(4f_\pi)^4} \int_{-1}^1
\!\!dy \int_{-1}^1 \!\!dz\, {yz \,\theta(y^2+z^2-1) \over  |yz|\sqrt{y^2+z^2-1}}
\bigg\{{2 y^4\,\theta(y)\theta(z)\over u(1+4u^2y^2)}\nonumber \\ && \times  
\Big[4u^2z^2-\ln(1+4u^2z^2)\Big]+ \int_0^u \!\! dx\,{5x^4 s^3s't^3  t'(y^2+z^2-1) 
\over  4u^{10}(1+s^2)(1+t^2)}\bigg\}\,, \end{eqnarray}
with the auxiliary function $t= x z +\sqrt{u^2-x^2+x^2z^2}$ and its partial
derivative $t' = u\, \partial t/\partial u$. For the numerical evaluation of the
$dy\,dz$-double integrals in eqs.(18,19,20) it is advantageous to first
antisymmetrize the integrands in $y$ and $z$ and then to substitute $z=
\sqrt{y^2\zeta^2+1-y^2}$. This way the integration region becomes equal to the
unit-square $0<y,\zeta<1$. 
\subsection{Irreducible two-pion exchange}
At next order in the small momentum expansion comes the irreducible
$2\pi$-exchange including (also) intermediate $\Delta$-isobar excitations. We 
employ a (subtracted) spectral-function representation of the $\pi N\Delta
$-loops and find the following non-vanishing (two-body) contribution:   
\begin{eqnarray} G_J(\rho) &=& {3\over 16\pi} \int_{2m_\pi}^\infty\!\! d\mu\, \Bigg\{
{\rm Im}(V_C-W_C)\bigg[{\mu \over 4k_f^6}(\mu^2+2k_f^2)\ln\bigg(1+{4k_f^2 
\over \mu^2}\bigg)-{\mu  \over k_f^4}-{4\over 3 \mu^3}\bigg] \nonumber \\ && +
{\rm Im}(V_T-W_T)\bigg[ {\mu  \over k_f^2} -{4 \over 3\mu}+{\mu^3 \over 2k_f^4} 
-{\mu^3 \over 8k_f^6}(\mu^2+4k_f^2)\ln\bigg(1+{4k_f^2 \over \mu^2}\bigg)
\bigg] \Bigg\}\,. \end{eqnarray} 
The imaginary parts Im$V_C$, Im$W_C$, Im$V_T$ and Im$W_T$ of the isoscalar and
isovector central and tensor NN-amplitudes due to $2\pi$-exchange with single
and double $\Delta$-excitation can be found in section 3 of 
ref.\cite{spectral}. The additional contributions from the irreducible 
$2\pi$-exchange with only nucleon intermediate states are accounted for by
inserting into eq.(21) the imaginary parts:
\begin{equation} {\rm Im}W_C= {\sqrt{\mu^2-4m_\pi^2} \over 3\pi 
\mu (4f_\pi)^4} \bigg[ 4m_\pi^2(1+4g_A^2-5g_A^4) +\mu^2(23g_A^4-10g_A^2-1) + 
{48 g_A^4 m_\pi^4 \over \mu^2-4m_\pi^2} \bigg] \,, \end{equation}
\begin{equation} {\rm Im}V_T= - {6 g_A^4 \sqrt{\mu^2-4m_\pi^2} \over 
\pi  \mu (4f_\pi)^4}\,. \end{equation}
At leading order the irreducible $2\pi$-exchange generates no spin-orbit
NN-interaction. It emerges first as a relativistic $1/M$-correction. In order to
see the size of such relativistic effects we have evaluated the energy 
density functional with a two-body interaction composed of the (isoscalar and
isovector) spin-orbit NN-amplitudes $V_{\rm SO}$ and $W_{\rm SO}$ written in
eqs.(22,23) of ref.\cite{nnpap}. We find with it the following contribution to 
the isovector spin-orbit coupling strength: 
\begin{eqnarray} G_{so}(\rho) &=&{g_A^2 m_\pi\over\pi M(4f_\pi)^4}\bigg\{{4-g_A^2 
\over 15u^4} \Big[\ln(1+u^2) - u^2\Big] +{3\over 10}(11g_A^2-4) \nonumber \\ && 
+\bigg[ {4-10g_A^2 \over 3u}+{4u \over 5}(1-4g_A^2) \bigg] \arctan u \bigg\}\,, 
\end{eqnarray}
which has been subtracted at $\rho=0$ in order to eliminate (regularization
dependent) short-distance components. As a consequence of that subtraction only
the Fock terms are included in the expressions in eqs.(21,24).
\subsection{Three-body diagrams with $\Delta$-excitation}
\begin{figure}
\begin{center}
\includegraphics[scale=1.0,clip]{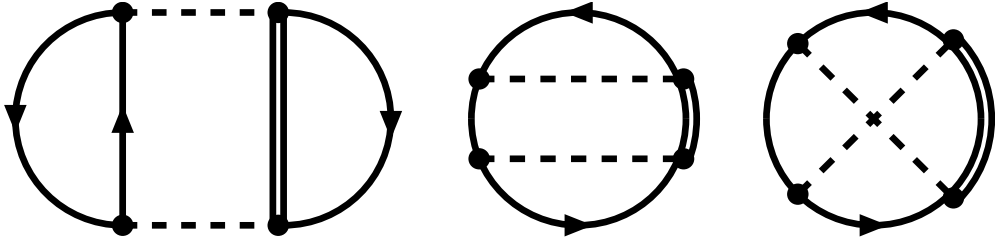}
\end{center}
\vspace{-0.2cm}
\caption{Hartree and Fock three-body diagrams related to $2\pi$-exchange with 
virtual $\Delta$-isobar excitation (''Fujita-Miyazawa \cite{fujita} 
mechanism'').} 
\end{figure}

The Pauli-blocking correction to the $2\pi$-exchange with single 
$\Delta$-excitation is equivalent to the contribution of a (genuine)
three-nucleon force. In fact, one is dealing here with the same three-nucleon 
interaction as originally introduced by Fujita and Miyazawa \cite{fujita}.
The pertinent Hartree and Fock three-body diagrams related to $2\pi$-exchange
with virtual $\Delta$-excitation are shown in Fig.\,2. Returning to the medium 
insertion written in eq.(4) we find from the left three-body Hartree diagram 
in Fig.\,2 the following (non-vanishing) contribution:
\begin{equation}   G_J(\rho) ={ g_A^4 m_\pi \over \pi^2\Delta(4f_\pi)^4}
\bigg\{{6\over u}-4u-{8u \over 1+4u^2} - {3\over 2u^3} \ln(1+4u^2)\bigg\}
\,,\end{equation}
with $\Delta = 293\,$MeV the delta-nucleon mass splitting. We have used the 
value $3/\sqrt{2}$ for the ratio between the $\pi N\Delta$- and 
$\pi NN$-coupling constants. The  vanishing of contributions to $G_d(\rho)$ and
$G_{so}(\rho)$ has the following reason. The pertinent isospin trace generates 
an expression that is odd under exchange of the momenta $\vec p_1$ and  $\vec
p_2$ attached to the (upward and downward running) nucleon lines of the (left) 
closed nucleon ring. The remaining factor from the two-pion exchange 
interaction is however even (under  $\vec p_1 \leftrightarrow \vec p_2$) and
so the whole expression integrates to zero.

The three-body effects on the energy density functional ${\cal E}_{\rm  iv}[\rho_p,
\rho_n,\vec J_p,\vec J_n]$  are completed by the contributions from the
central and right Fock diagrams in Fig.\,2, which read:
\begin{eqnarray}G_d(\rho) &=& { g_A^4 m_\pi\over 3\pi^2\Delta(8 f_\pi)^4}\Bigg\{ 
{112u \over 1+4u^2}-16u -{20\over u}+{30\over u^3}\nonumber\\ && + \bigg(
{64u \over 1+4u^2}-{14\over u}-{20\over u^3} -{15 \over u^5}\bigg)\ln(1+4u^2) 
\nonumber \\ && +{5\over 8 u^7}(3+10u^2+8u^4)\ln^2(1+4u^2) \Bigg\}\,,
\end{eqnarray}
\begin{eqnarray}G_{so}(\rho) &=& { 7g_A^4 m_\pi\over \pi^2\Delta(8 f_\pi)^4}\Bigg\{ 
{64u \over 9}-{4\over 3u}-{7\over 3u^3}-{4\over u^5}-{5\over 4u^7} \nonumber
\\ && + \bigg({5\over 8u^9}+ {13\over 4u^7}+ {13\over 3u^5} + {2\over u^3} - 
{8\over 3u} \bigg)\ln(1+4u^2) \nonumber \\ && -{1\over 64 u^{11}}
(64u^6+80u^4+36u^2+5) \ln^2(1+4u^2) \Bigg\}   \,,\end{eqnarray}
\begin{eqnarray}G_J(\rho) &=& { g_A^4 m_\pi\over \pi^2 \Delta(8 f_\pi u)^4}\Bigg\{ 
\bigg[{107u^4\over 2}+{217u^2\over 4}-{377\over 4} +{8\over u^2} \ln(1+4u^2)
\bigg] \arctan 2u \nonumber \\ && +\bigg({3\over 8u^5} +{5\over  4u^3}+{737
\over 64 u}-{3547 u\over 192}+{81 u^3\over 16}\bigg) \ln(1+4u^2) -{3\over 4u^3}
-{1\over u} \nonumber \\ &&+{787u \over 16}-{3935u^3\over 48}-{1696 u^5 \over
15}  - {3+16u^2 +144u^4\over 64u^7}\ln^2(1+4u^2) \nonumber \\ &&
 +\int_0^u\!\!dx\,\bigg\{{3L^2 \over 8u^2}\bigg[ {42\over   x^2}(1+u^2)^4(3u^2-1)
-{35\over x^4}(1+u^2)^6+(1+u^2)^2\nonumber \\ && \times (94u^2-145-257u^4) + 4x^2
(117u^6+35u^4-145u^2-63) -103x^8 \nonumber \\ && +x^4(182u^2-165-597u^4) +2x^6
(199u^2-77)\bigg]+{L \over 4u}\bigg[{105\over  x^4}(1+u^2)^5 \nonumber \\ && 
+{7\over x^2}(1+u^2)^3(3-49u^2) +18(1+u^2)(37u^4-4u^2+23) \bigg]
\nonumber \\&& -{105\over  8x^4}(1+u^2)^4 +{7\over  2x^2}(1+u^2)^2(3+11u^2) 
\bigg\}\Bigg\}   \,,\end{eqnarray}
with $L(x,u)$ defined in eq.(17). A good check of all formulas collected in 
this section is provided by their Taylor series expansion in $k_f$. Despite the
superficial opposite appearance the leading term in the $k_f$-expansion is  
always a non-negative power of $k_f$ (which is higher for three-body 
contributions than for two-body contributions).
\section{Results and discussion}
In this section we present and discuss our numerical results obtained by
summing the series of contributions given in section 3. The physical input
parameters are: $g_A=1.3$ (nucleon axial vector coupling constant), $f_\pi = 
92.4\,$MeV (pion decay constant), $m_\pi = 135\,$MeV (neutral pion mass) and 
$M=939\,$MeV (nucleon mass). We recall that with these physical parameters and 
a few adjustable short-distance couplings the nuclear matter equation of state 
$\bar E(\rho)$ and many other nuclear matter properties \cite{deltamat} can be 
well described by the chiral pion-nucleon dynamics treated up to three-loop 
order. 

\begin{figure}
\begin{center}
\includegraphics[scale=0.55,clip]{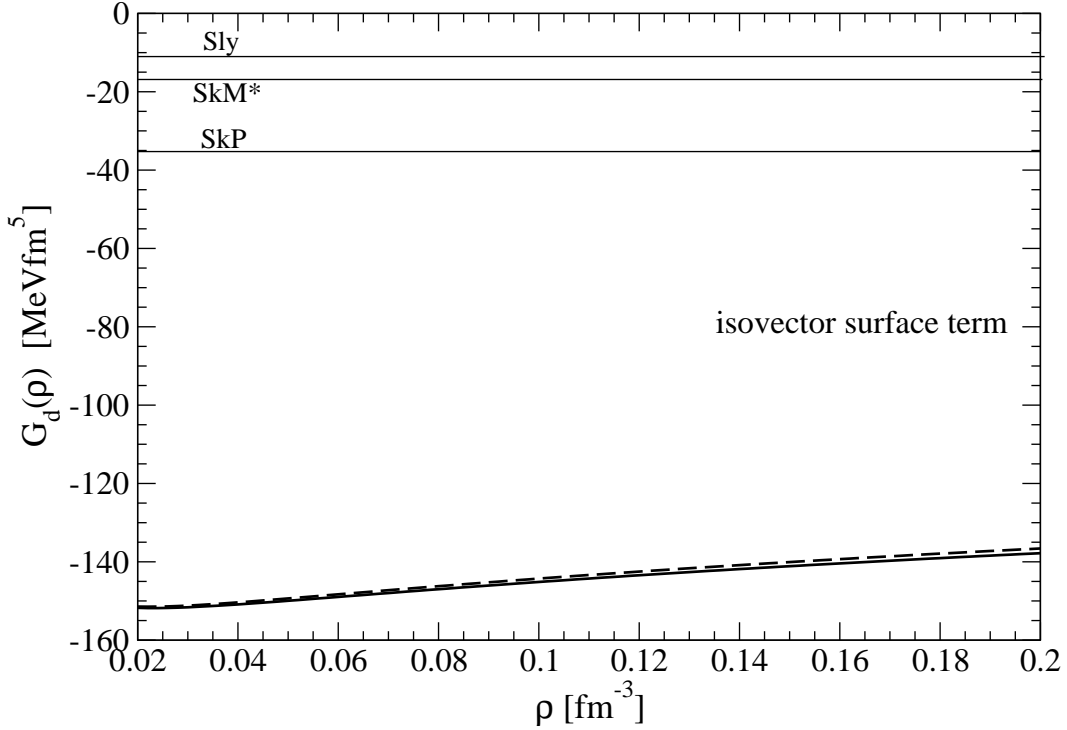}
\end{center}
\vspace{-0.2cm}
\caption{Strength function $G_d(\rho)$ of the isovector surface term 
$(\vec \nabla \rho_p-\vec \nabla \rho_n)^2$ in the nuclear energy density 
functional versus the nucleon density $\rho=2k_f^3/3\pi^2$. Dashed line: 
Iterated $1\pi$-exchange only. Full line: $2\pi$-exchange and associated 
three-body contributions added.}
\end{figure}

Fig.\,3 shows the strength function $G_d(\rho)$ belonging to the isovector 
surface term $(\vec \nabla \rho_p-\vec \nabla \rho_n)^2$ plotted versus the 
nucleon density $\rho=2k_f^3/3\pi^2$. One observes that the leading result due 
to the iterated $1\pi$-exchange (eqs.(6,8,11,14,18) shown separately by the 
dashed line) is almost not changed by the inclusion of three-body contribution 
eq.(26) related to $2\pi$-exchange with virtual $\Delta$-excitation. Moreover,
the density dependence of  $G_d(\rho)$ is rather weak in the entire region $0
< \rho <0.2\,{\rm fm}^{-3}$. The horizontal lines in Fig.\,3 correspond to the 
(constant) values $G_d^{(\rm Sk)} = -[3t_1(2x_1+1)+t_2(2x_2+1)]/64$ of three
phenomenological Skyrme forces SkM$^*$ \cite{skmstar}, SkP \cite{skp} and Sly 
\cite{sly}. These values are of the same negative sign, but considerably
smaller in magnitude than our parameter-free prediction resulting from the
long-range pion-exchange dynamics in the nuclear medium. It remains to be seen 
how well these (predicted) much larger negative values of $G_d(\rho)$, which
energetically favor large differences in the density-gradients of protons and
neutrons, perform in actual nuclear structure calculations at large neutron 
excess.

\begin{figure}
\begin{center}
\includegraphics[scale=0.55,clip]{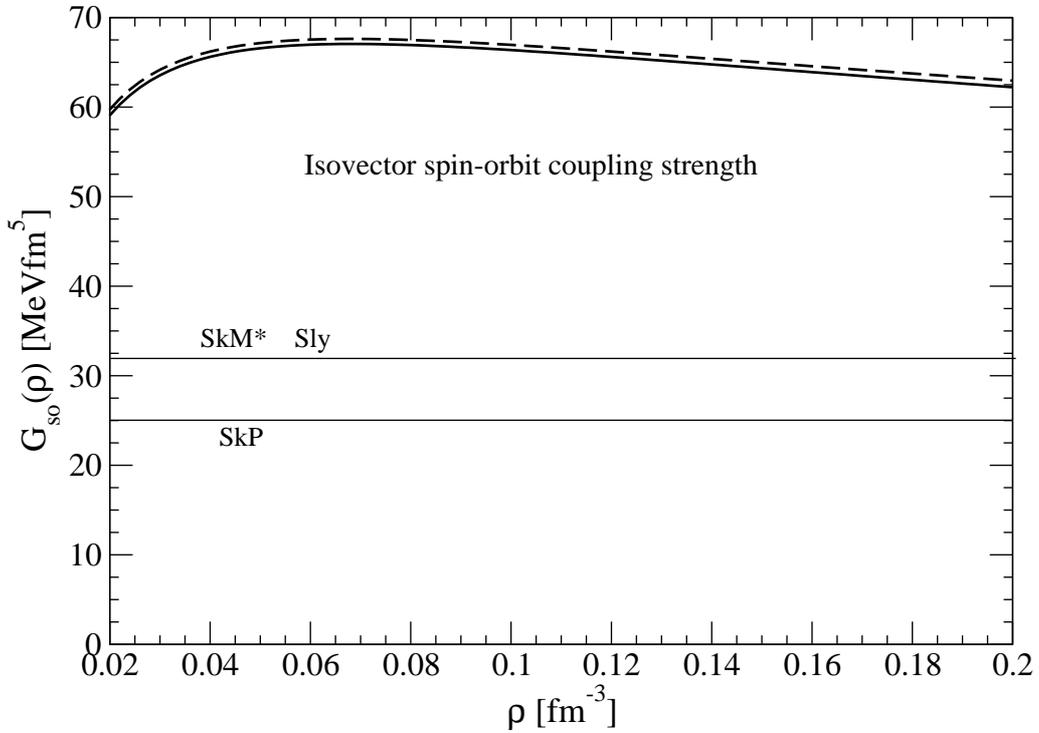}
\end{center}
\vspace{-0.2cm}
\caption{Strength function $G_{so}(\rho)$ of the isovector spin-orbit
coupling term $(\vec \nabla \rho_p-\vec \nabla\rho_n)\cdot(\vec J_p-\vec J_n)$ 
in the nuclear energy density  functional versus the nucleon density
$\rho=2k_f^3/3\pi^2$. Dashed line: Iterated $1\pi$-exchange only. Full 
line: $2\pi$-exchange and three-body contributions added.}
\end{figure}

Of particular interest is the strength function $G_{so}(\rho)$ of the isovector 
spin-orbit coupling term $(\vec \nabla \rho_p-\vec \nabla\rho_n)\cdot(\vec J_p
-\vec J_n)$ as provided by the explicit pion-exchange dynamics. As one can see
from Fig.\,4 the leading result due to the iterated $1\pi$-exchange 
(eqs.(7,9,12,15,19) shown separately by the dashed line) is again not changed
by the inclusion of the $2\pi$-exchange and associated three-body 
contributions eqs.(24,27). This feature is markedly different from the 
isoscalar spin-orbit coupling strength $F_{so}(\rho)$ where an almost complete 
cancellation between these two components has occurred around $\rho_0/2 = 0.08
\,{\rm fm}^{-3}$ (see Fig.\,5 in ref.\cite{energfun}). The basic reason for this
totally different behavior is the absence of a strong isovector three-body
spin-orbit coupling arising through the Fujita-Miyazawa mechanism (i.e. from 
the dominant three-body Hartree diagram in Fig.\,2). There is of course in 
addition the short-range spin-orbit interaction which has to account 
practically for the full strength $F_{so}^{(\rm  emp)}\simeq 90\,$MeVfm$^5$ in the
isoscalar channel and which contributes also in the isovector channel with a
reduced weight $1/3$. Combined with our result for $G_{so}(\rho)$ from the 
long-range pion-exchange dynamics one would then have a situation where the 
isoscalar and isovector spin-orbit coupling strengths are about equally 
strong. As discussed in ref.\cite{isoorbit} the isotope shifts of the charge 
radii in the Pb region (i.e. around the shell-closure $N=126$) can provide a 
sensitive test for strength of the isovector spin-orbit coupling. However, 
according to their limited analysis (in the vicinity of the stability line) no
definite choice could be made between several density-dependences of the 
neutron spin-orbit potential: $\sim \rho_p +2\rho_n$ for the Skyrme force, 
$\sim \rho_p +\rho_n$ for the relativistic mean-field model, or $\sim \rho_p$
for the generalized functional SkI4 \cite{isoorbit}. It remains to be seen
whether the proportionality of the neutron spin-orbit potential to the
gradient of the neutron density $\vec \nabla \rho_n$ as suggested by the 
present calculation leads to results which are in accordance with the existing
precise experimental data. 
 
\begin{figure}
\begin{center}
\includegraphics[scale=0.55,clip]{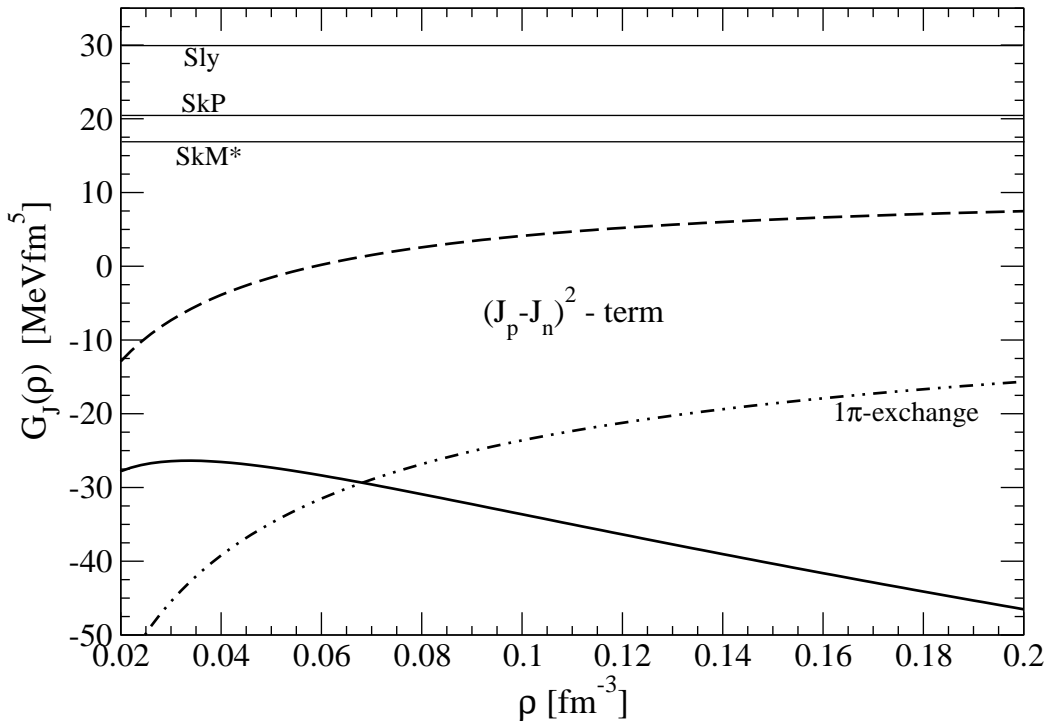}
\end{center}
\vspace{-0.2cm}
\caption{Strength function $G_J(\rho)$ belonging to the squared isovector
spin-orbit density $(\vec J_p-\vec J_n)^2$ in the nuclear energy density 
functional versus the nucleon density $\rho=2k_f^3/3\pi^2$. Dashed line: 
$1\pi$- and  iterated $1\pi$-exchange only. Full line: $2\pi$-exchange and 
three-body  contributions added.}
\end{figure}

Finally, we show in Fig.\,5 the strength function $G_J(\rho)$ belonging the 
squared isovector spin-orbit density $(\vec J_p-\vec J_n)^2$ in the nuclear 
energy density functional as a function of the nuclear density $\rho$. One 
observes that the inclusion of the subleading $2\pi$-exchange (two- and 
three-body) contributions eqs.(21,25,28) significantly affects the outcome for 
$G_J(\rho)$. For better orientation, we reproduce also by the dashed-dotted 
line in Fig.\,5 the leading contribution from the $1\pi$-exchange Fock diagram 
alone (see eq.(5)). Interestingly, in the density region around $\rho_0/2 = 
0.08\,{\rm fm}^{-3}$ the $1\pi$-exchange approximation and the complete result 
(full line) agree roughly with each other. The constant values $G_J^{(\rm Sk)}= 
(t_1-t_2)/32$ from phenomenological Skyrme forces \cite{skmstar,skp,sly} are 
of similar magnitude but opposite in sign. Note that the strength function 
$G_J(\rho)$ comprises in particular the non-local Fock contributions from 
tensor forces, whose long-range isovector component is determined by the 
$1\pi$-exchange. This correspondence gives a rough understanding of the 
features visible in Fig.\,5.  Another interesting side effect of the $(\vec 
J_p-\vec J_p)^2$ term in the nuclear energy density functional is that it gives 
rise to an extra spin-orbit mean-field $2G_J(\rho)\, (\vec J_p- \vec J_n)$ in 
addition to the ''normal'' one, $G_{so}(\rho)\, (\vec \nabla \rho_p -\vec \nabla 
\rho_n)$. It would be interesting to investigate the role of this additional 
(nucleus-dependent)  spin-orbit mean-field together with the density-dependence 
of $G_J(\rho)$ as predicted by in medium chiral perturbation theory. 

In summary, we have used the improved density-matrix expansion of 
ref.\cite{dmeimprov} to calculate the strength functions of the isovector 
surface and spin-orbit terms in the nuclear energy density functional as 
provided by the long-range pion-exchange dynamics in the nuclear medium. These 
predictions together with the ones in ref.\cite{energfun} for the isoscalar 
strength functions should be examined and explored in nuclear structure 
calculations at small and large neutron excess.


\begin{thebibliography}{99}
\bibitem{reinhard} M. Bender, P.H. Heenen and P.G. Reinhard, {\it Rev. Mod.
Phys.} {\bf  75} (2003) 121;\\ J.R. Stone  and P.G. Reinhard, {\it Prog. 
Part. Nucl. Phys.} {\bf  58} (2007) 587.\vs
\bibitem{sk3} M. Beiner, H. Flocard, N. Van Giai and P. Quentin, \textit{Nucl. 
Phys.} \textbf{A238} (1975) 29.\vs
\bibitem{skmstar} J. Bartel, P. Quentin, M. Brack, C. Guet and H.B. Hakansson,
\textit{Nucl. Phys.} \textbf{A386} (1982) 79.\vs
\bibitem{skp} J. Dobaczewski, H. Flocard and J. Treiner, \textit{Nucl. Phys.} 
\textbf{A422} (1984) 103.\vs 
\bibitem{sly} E. Chabanat, P. Bonche, P. Haensel, J. Meyer and R. Schaeffer,
\textit{Nucl. Phys.} \textbf{A627} (1997) 710; \textbf{A635} (1998) 231; 
and refs. therein.\vs
\bibitem{pearson} S. Goriely, M. Samyn, M. Bender and J.M. Pearson, {\it
Phys. Rev.}  {\bf C68} (2003) 054325;\\ M. Samyn,  S. Goriely, M. Bender and 
J.M. Pearson, {\it Phys. Rev.}  {\bf C70} (2004) 044309; \\ N. Chamel, 
S. Goriely and J.M. Pearson, {\it  Nucl. Phys.} {\bf A812} (2008) 72.\vs
\bibitem{walecka} B.D. Serot and J.D. Walecka, {\it Int. J. Mod. Phys.} {\bf 
E6} (1997) 515; and refs. therein.\vs 
\bibitem{ringreview} P. Ring, Lecture Notes in Physics, Vol.581, 
Springer-Verlag, Berlin, 2001, p. 195; and refs. therein.\vs
\bibitem{finelli} P. Finelli, N. Kaiser, D. Vretenar and W. Weise, 
\textit{Nucl. Phys.} \textbf{A770} (2006) 1.\vs
\bibitem{lesinski} T. Lesinski, T. Duguet, K. Bennaceur and J. Meyer, 
\textit{Eur. Phys. J.} \textbf{A40} (2009) 121.\vs 
\bibitem{drut} J.E. Drut, R.J. Furnstahl and L. Platter, {\it Prog. Part. 
Nucl. Phys.} {\bf 64} (2010) 120; nucl-th/0906.1463.\vs
\bibitem{platter} S.K. Bogner, R.J. Furnstahl and L. Platter, \textit{Eur. 
Phys. J.} \textbf{A39} (2009) 219.\vs
\bibitem{achim} S.K. Bogner, R.J. Furnstahl, A. Nogga and A. Schwenk, {\it Nucl.
Phys.} {\bf A763} (2005) 59; nucl-th/0903.3366.\vs
\bibitem{roth} R. Roth, P. Papakonstantinou, N. Paar, H. Hergert, T. Neff and
H. Feldmeier, {\it Phys. Rev.} {\bf C73} (2006) 044312.\vs
\bibitem{negele} J.W. Negele and D. Vautherin, {\it Phys. Rev.} {\bf C5} (1972)
1472.\vs 
\bibitem{dmeimprov} B. Gebremariam, T. Duguet and S.K. Bogner, 
nucl-th/0910.4979.\vs 
\bibitem{energfun} N. Kaiser and W. Weise, \textit{Nucl. Phys.} \textbf{A} 
(2010) in print; nucl-th/0912.3207.\vs
\bibitem{short} N. Kaiser,  {\it Phys. Rev.} {\bf C70} (2004) 034307.\vs
\bibitem{tueb} O. Plohl and C. Fuchs,  {\it Phys. Rev.} {\bf C74} (2006) 
034325.\vs
\bibitem{efun} N. Kaiser, S. Fritsch and W. Weise, \textit{Nucl. Phys.} 
\textbf{A724} (2003) 47.\vs
\bibitem{spectral} N. Kaiser, S. Gerstend\"orfer and W. Weise, \textit{Nucl. 
Phys.} \textbf{A637} (1998) 395.\vs
\bibitem{nnpap} N. Kaiser, R. Brockmann and W. Weise, \textit{Nucl. Phys.} 
\textbf{A625} (1997) 758.\vs
\bibitem{fujita} J. Fujita and H. Miyawawa,  {\it Prog. Theor. Phys.} {\bf 17} 
(1957) 360; 366.\vs
\bibitem{deltamat} S. Fritsch, N. Kaiser and W. Weise, \textit{Nucl. Phys.} 
\textbf{A750} (2005) 259.\vs
\bibitem{isoorbit} P.G. Reinhart and H. Flocard, \textit{Nucl. Phys.} 
\textbf{A584} (1995) 467.\vs
\end{thebibliography}
\end{document}